# The peloton superorganism and protocooperative behavior

Hugh Trenchard

805 647 Michigan Street, Victoria, BC V8V 1S9, Canada

**Abstract**

A theoretical framework for protocooperative behavior in pelotons (groups of cyclists) is proposed. A threshold between cooperative and free-riding behaviors in pelotons is modeled, together comprising protocooperative behavior (unlike protocooperation). Protocooperative behavior is a function of: 1. two or more cyclists coupled by drafting benefit; 2. current power output or speed; and 3. cyclists maximal sustainable outputs (MSO [1]). Characteristics of protocooperative behavior include: 1. relatively low speed phase in which cyclists naturally pass each other and share highest-cost front position; and 2. free-riding phase in which cyclists maintain speeds of those ahead, but cannot pass. Threshold for protocooperative behavior is equivalent to coefficient of drafting ($d$), below which cooperative behavior occurs; above which free-riding occurs up to a second threshold when coupled cyclists diverge. Range of cyclists' MSOs in free-riding phase is equivalent to the energy savings benefit of drafting ($1-d$). When driven to maximal speeds, groups tend to sort such that their MSO ranges equal the free-riding range ($1-d$). These behaviors are also hypothesized to emerge in biological systems involving energy savings mechanisms. Further, the tension between intra-group cooperation and inter-group competition is consistent with superorganism properties.

## 1. Introduction

"No man can be a good citizen unless he has a wage more than sufficient to cover the bare cost of living, and hours of labor short enough so that after his day's work is done he will have time and energy to bear his share in the management of the community, to help in carrying the general load" [1].

Theodore Roosevelt's quote well captures the notion that cooperation can occur only if individuals are not engaged in activity that require their maximum effort; that they must have some luxury or spare energetic resources to contribute to a cooperative effort. This is the essential conclusion of the work here.

Broadly, cooperators incur costs for the benefit of others; defectors have no costs and provide no benefits [2]. Because natural selection favors defectors [2], how cooperation emerges and proliferates in natural conditions has long been a puzzle of evolutionary biologists. In [2] Nowak outlines five basic mechanisms for the evolutionary selection of cooperators, generally studied using game theoretical models: kin selection (genetic relations are selected for), direct reciprocity (returning favor), indirect reciprocity (enhanced reputation through cooperation), network reciprocity (clusters can favor cooperators in certain circumstances), and group selection. Group selection posits that cooperators may be favored in certain conditions of competition between groups [2].

---

[1] *Abbreviations:* **PCR**, peloton convergence ratio; given by Eq. (2); **MSO**, maximal sustainable output; the power output sustainable by a cyclist (or any organism) until exhaustion at different physiological thresholds [28] the range of power outputs sustained for 14 cyclists during 200 m time trials as reported in [28] is the range of MSOs used for the peloton algorithm here; **PBT**, protocooperative behavior threshold; **RAP**, random acceleration parameter; random acceleration algorithm as added to the general algorithm in [15].

Game theoretical approaches to the study of cooperation, originating perhaps with [3,4], consider games such as the prisoner's dilemma and the snowdrift game to study cooperation between unrelated individuals [5,6]. More recent extensions to cooperation theory approaches include graph and network analysis [7–10].

Cooperative dynamics have been explored in bicycle racing. In [11] the authors study cyclists' cooperative behavior over a normally distributed range of cyclists' power outputs. Conclusions in [11] were that riders with above average strength fared better as cooperators, and those with below average strength fared better as defectors, and that strategies of stronger riders also affected the results of their team-mates. This is consistent with the model and theory presented here, but in [11] cooperative probabilities were introduced as a model parameter to simulate human-based strategies, while here we find cooperative behavior emerges as a self-organized consequence of the model.

In [12], the probability of winning was modeled as a function of group size and energetic costs. The main finding was the optimal group size for breakaway success, based on data from [13]. In [14], a mathematical model of relative group speeds was presented based on a number of simplifying assumptions, including that riders were identical; riders shared the lead evenly; they neither accelerated nor decelerated; maintained the same power output that they would if riding alone. All of these factors are variables in the model here, allowing us to capture a broader range of behaviors.

Here we develop an earlier model of peloton dynamics [15], and advance peloton theory into the realm of evolutionary biology and the origins of cooperation. We explore conditions for group divisions and the effects of cyclists' physiological maximal sustainable outputs (MSO) on self-organized cooperative behavior. We argue that cyclists' non-volitional self-organized cooperative and free-riding behavior, and the threshold between them, represents a form of primitive cooperative behavior that is not, to our knowledge, described elsewhere as proposed.

We propose that "protocooperative behavior" – a different phenomenon from proto-cooperation as a form of mutualism [16] – self-organizes when two or more cyclists are coupled by the energy savings mechanism of drafting which occurs by riding behind others in zones of reduced air-resistance [17], and when cyclists expend energy above and below a passing capacity threshold, which we refer to as the "protocooperative behavior threshold". We hypothesize that this behavior generalizes to any biological system coupled by an energy savings mechanism.

Further, we argue that the characteristics of protocooperative behavior are analogous and consistent with behavior that characterizes superorganisms, as described in [18]. The conclusions in [18] are that an individual's evolutionary stable investment in within-group cooperation versus within-group competition correlate to genetic relatedness, group size, the number of competing groups, and between-group relatedness.

## 1.1. The drafting quantities and the divergence threshold

Except for a proportionately small number of riders in non-drafting positions at the front of the peloton, cyclists in a peloton ride behind other cyclists in a zone of reduced air-pressure, known as drafting. Empirical studies are somewhat inconsistent in the energy savings or power output reductions (numerically equivalent [19]): Kyle [20] shows ~29% power reduction at 24 km/h, ~31% power reduction at 32 km/h, ~33% power reduction at 40 km/h, and ~34% at 48 km/h; McCole et al. [17] shows reductions in metabolic costs of ~18% (±11) at 32 km/h, ~28% (±10) at 37 km/h, and ~26% (±8) at 40 km/h; Zdravkovich et al. [21] reported 37% and 35% wind-resistance reductions at ~30 km/h for a specific on-bicycle cyclist posture; Broker et al. found an ~29% power decrease for the second rider in a four-person team pursuit, and an additional 7% for the third rider [22]; Blocken et al. reported drag reductions for a following cyclist to be 27.1%, 23.1% and 13.8% depending on posture at 0.1 m wheel separation at 20 m/s using airflow only (no wheel and leg movement) and without considering bicycle drag (cyclist bodies only) [23].

The drafting coefficient is described by Olds [14]

$$d = 0.62 - 0.0104 d_w + 0.0452_w^2. \qquad (1)$$

Olds' Eq. (1) is a ratio of the wind resistance of a drafting rider to the wind resistance of a leading, non-drafting rider. It accounts for variations in *d* as distance varies between a leading cyclist's rear wheel and a following cyclist's front wheel. Olds derived his equation from Kyle [20] who found a 38% reduction in wind resistance for a drafting cyclist at 40 km/h and applied it as a constant for speeds between 24 km/h and 56 km/h. Olds introduces Eq. (1) as a correction factor to the power output required to overcome wind resistance [14]. The value 1-*d* is therefore equivalent to the energy saved by the drafting rider [19], which is approximately 38%, given optimal wheel spacing and normalized for other lesser factors that may affect the power output of the drafting rider [24].

Olds' Eq. (1), however, does not well account for drafting as a function of speed variations (effective wind speed), particularly at speeds less than 32 km/h (8.9 m/s) [20], and where there is negligible drafting benefit at < 16 km/h [25]. Eq. (1) does not appear to have been updated in the literature to account for variations in cyclists' travelling speed (i.e. effective wind-speed and drag), which variations have significant effects as cyclists approach their maximal outputs when cycling uphill at speeds < 32 km/h. Thus while drafting models are incomplete to account for wheel separation, effective wind-angle, and sideways distance [26], so also are drafting models that account for effective wind-speed of less than 32 km/h (8.9 m/s). In turn, researchers continue to cite and apply Eq. (1) [19,26,27]. Similarly, in this study, we apply Eq. (1) and the coefficient of 0.62 for *d* and 0.38 for 1-*d*, assuming a constant optimal wheel spacing at all speeds. Future research should account for the effects of variations in *d*.

The equation that describes coupling behavior between cyclists is, as shown in [15,28],

$$\text{PCR} = \frac{P_{front} - [P_{front} * (1-d)]}{\text{MSO}_{follow}} \quad \text{and simplified:} \quad \text{PCR} = \frac{P_{front} * d}{\text{MSO}_{follow}}. \qquad (2)$$

PCR is the "peloton convergence ratio", describing two coupled riders: the non-drafting front-rider sets the pace; the follower obtains the drafting benefit of reduced power output at the same speed as the front-rider. Two-cyclist coupling generalizes to many-rider interactions. All drafting cyclists are coupled both to the rider immediately ahead who provides the drafting benefit for the follower, and to a single or small number of non-drafting cyclist at the front of the peloton. Here we refer to a "front-rider" as a non-drafting cyclist who sets the pace; a "leader" is a cyclist immediately in front of a drafting rider, but who herself may be drafting behind other riders.

"$P_{front}$" is the power output of the front-rider as she sets the pace within the coupled system; $d$ is the drafting coefficient from Eq. (1).

"$MSO_{follow}$" is the maximal sustainable power output of the follower. If PCR > 1 between two cyclists, the follower cannot sustain the speed set by the leader and must therefore decelerate to a speed less than or equal to the speed equivalent of the cyclist's maximum capacity (MSO), as shown in Eqs. (3)–(6) [15].

Following cyclists are forced to decelerate when their outputs are effectively driven over their MSOs by stronger front riders whose power requirements exceed the drafters' MSOs. In this study we are concerned primarily with deceleration processes by physiological necessity rather than by intention and choice, although we can simulate volitional deceleration by manually adjusting (reducing) the speeds set by the front rider (effective peloton speed). Self-organized (non-volitional) deceleration is modeled in [15], and summarized as follows:

First rearrange (2) to obtain the front-rider's power output ("$P_{front}$"):

$$P_{front} = \frac{MSO_{follow} * PCR}{d}. \tag{3}$$

Then convert "$P_{front}$" to velocity ($V_{front}$) applying power output relationships and parameters as in [15, Appendix A]. A (weaker) follower must decelerate to a speed less than or equal to the speed that corresponds to PCR = 1, so we obtain the power output corresponding to her MSO, when drafting (i.e., PCR = 1):

$$P_{threshold} = \frac{MSO_{follow}}{d}. \tag{4}$$

This is Eq. (3) with the assumption PCR = 1. "$P_{threshold}$" thus represents power output reduction due to drafting when PCR = 1. The expression $P_{front} - P_{threshold}$ obtains the follower's speed corresponding to $MSO_{follow}$. Convert "$P_{threshold}$" to an equivalent speed ("$V_{threshold}$") using the same power-speed relationships as before [15, Appendix A], we then find the difference between the speed set by the (stronger) front-rider and the (slower) speed which is the maximal speed available to the (weaker) rider:

$$V_{reduction} = V_{front} - V_{threshold}. \tag{5}$$

"$V_{reduction}$" is thus the deceleration magnitude for the following rider when the output required to maintain the speed of a front-rider corresponds to a power output that exceeds her MSO (PCR > 1). Finally, when cyclists decelerate to keep their output at or below MSO, they usually slow to an output below, but not exactly at the threshold (on an individually varying basis), so a small random magnitude of deceleration is added:

$$V_{reduction} = V(P_{front}) - V(P_{threshold}) + \Delta V. \tag{6}$$

In (7) "$V(P_{front})$" is a velocity "$V$" expressed as a function of a power "$P$", and is the noted small positive random individual deceleration quantity.

## 1.2. The protocooperative threshold

A following, drafting cyclist may be working at or below MSO. If she is working at MSO while drafting but conditions change such that the follower moves outside the optimal drafting position, then the follower must decelerate (assuming the leading rider sustains her speed). If the follower is below MSO while drafting but temporarily moves outside drafting range, she can increase power output to maintain the pace of the leader as long as she does not exceed MSO.

For a drafting cyclist to accelerate past a leading rider and assume the front position, the follower's maximum ($MSO_{follow}$) must exceed the power ($P_{front}$) required for the speed set by the front rider. This capacity to pass is fundamental to protocooperative behavior, since it allows cyclists to share time spent at the front of the peloton, where energy requirements are highest. Passing behavior is a property of a high-density peloton in which cyclists travel in a compact, roughly circular but fluid two-dimensional geometrical formation [28]. As cyclists' speeds increase, their ability to pass diminishes up to a threshold when, by drafting, weaker cyclists can sustain the speed set by a stronger rider (or small number of stronger riders) who sets the pace in the most energetically costly position due to highest aerodynamic drag, but whom weaker (or equally strong) riders cannot pass.

We propose the "protocooperative behavioral threshold" (PBT) as the power output (translated to speed) threshold, below which cyclists are capable of passing, and above which cyclists can sustain the speed of cyclists ahead but cannot pass, as shown in Figs. 1 and 2. In this situation, cyclists cannot pass because to do so would require followers to shift to a non-drafting position and accelerate, requiring greater power than their MSO; however, in this situation the following cyclist may sustain the speed of the rider ahead because of the reduced power requirement of drafting.

Accordingly, there is a range of outputs above PBT in which weaker (or equally strong) drafting cyclists cannot pass riders ahead, as shown in Fig. 1. This is the free-riding range or phase in which following riders, by physiological limitation, cannot contribute to the most costly front position. This phase has also been described as a stretched phase in which the compact formation elongates and cyclists align one behind the other in single file [28]. Because cyclists approach maximal sustainable capacities in this phase, they have no spare energetic resources to contribute to the most costly front position, analogous to the circumstance described in the quote by Theodore Roosevelt [1].

In this phase we assume the front rider does not decelerate to accommodate a following rider's passing attempt. PCR approaches 1 for the coupled cyclists in this condition, while cyclists diverge when PCR = 1. If a front rider does decelerate to allow the follower to pass, then the phase shifts to a passing phase, such that riders are below PBT and where cooperative (passing) behavior emerges. Phase behaviors are visually most obvious among a many-rider peloton, and mixed phase behavior can occur such that some riders engage in passing behavior below the PBT, while others are in a stretched phase, above the PBT.

The protocooperative behavior threshold is:

$$PCR_{pbt} = d \tag{7}$$

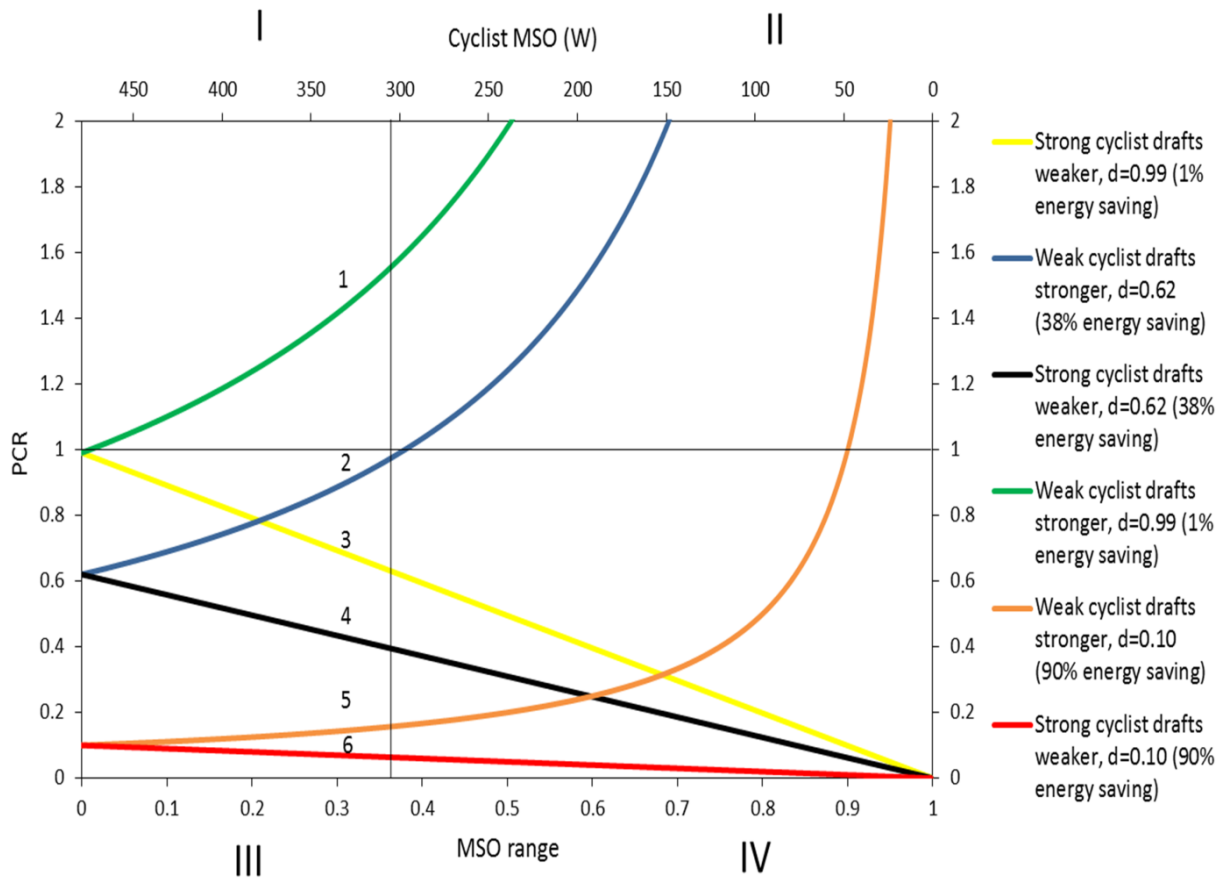

**Fig. 1.** Protocooperative behavior thresholds (PBT) for three *d* values (0.1, 0.62, 0.99), and relationships between MSO range, PCR, and cyclists' maximal power (MSO). Upper horizontal axis shows a range of MSOs from 0 W to 479 W; lower horizontal axis shows the range as proportions of 1. PBT is shown at convergence points between straight slopes indicating the PCRs for the passing phase (stronger rider drafts weaker who sets pace at MSO), and curves showing PCRs for the free-riding (non-passing) phase (weaker rider drafts stronger who sets pace at MSO), points PCR = 0.1, 0.62, and 0.99 (equivalent to *d*). Free-riding range is 1-*d*. Narrow horizontal line shows divergence threshold PCR = 1; narrow vertical line shows cyclist with MSO = 305 W, the lowest in the range 305–479 W.

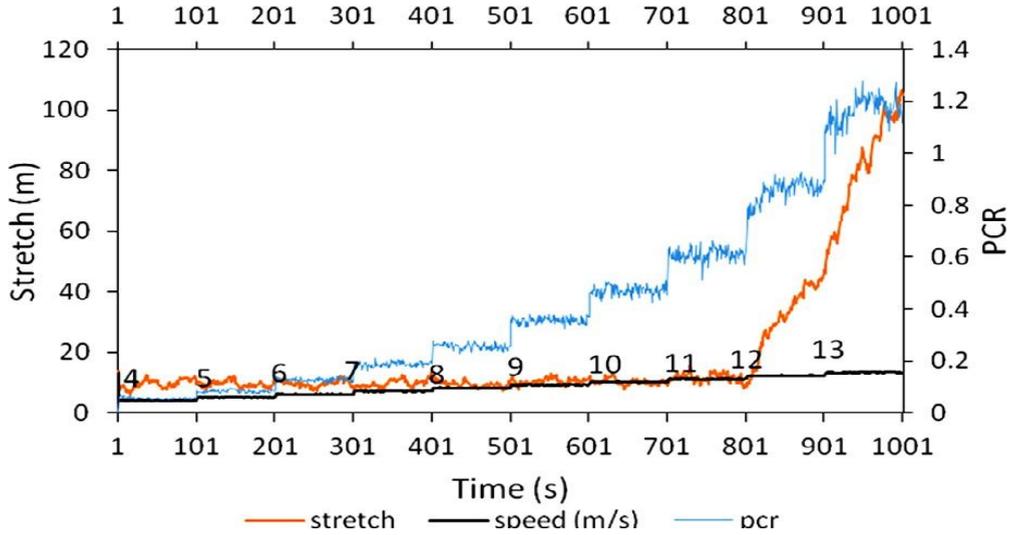

**Fig. 2.** Typical peloton stretch and PCR obtained for 50 simulated cyclists by incremental increases in collective speed by 1.0 m/s, from 4.0 m/s (14.4 km/h) to 13.0 m/s (46.8 km/h), each with a random acceleration parameter 0.0–2.0 m/s. Stretch is distance (m) from the rider at the front of the group to the rider farthest back. Between speeds 4.0 m/s and 12.0 m/s (43.2 km/h), stretch is small, indicating high density. A phase change occurs at ~12.0 m/s in which the group stretches markedly. The pre-transition mean PCR = 0.62 for the 50 cyclists, corresponding approximately to $d$.

$PCR_{pbt}$ is the PCR value that corresponds to the protocooperative behavioral threshold. Cyclists coupled at PCR < $d$ are capable of passing by acceleration since, when this condition is satisfied, the drafting cyclist's MSO always exceeds the required output of the front position, non-drafting rider. As indicated, the capacity to pass allows cyclists to share the most costly front-riding position.

## *1.3. The properties of protocooperative behavior in pelotons*

Fig. 1 models the properties of the PBT, using a range of cyclists' MSOs between 0 W and 479 W (the "model range"). MSO = 305 W is the low value in the range of MSOs of 14 female track cyclists observed for the study in [28], shown on the upper axis (the "sample range"). The MSOs were derived from 200 m sprint times for 14 female cyclists, assumed to closely approximate their maximal sustainable power outputs, obtained from publicly accessible competition results [28]:

    479  458  435  412  402  400  397  397  397  393  372  356  351  305

Range as proportions of 1 is shown on the lower axis ($MSO_{max} - MSO_{min}/MSO_{max}$). Three sets of slopes show PCRs for three $d$ quantities for the model range: $d$ = 0.1 (energy savings = 90%), $d$ = 0.99 (energy savings = 1%) showing extreme cases[2] of maximum drafting and minimal drafting coefficients; and $d = 0.62$ (energy savings = 38%) from Eq. (1) which is applied to simulation results, discussed in sections following.

---

[2] $d$ = 0.1 was selected and not $d$ = 0.01 since it is easier to see intersecting points on the plot.

Straight slopes (3, 4, 6) show PCRs (to max PCR = 2) assuming the strongest rider (479 W) is the drafting rider, and the range of MSO = 1 W to MSO = 479 W are weaker (or equally strong) riders in the front position. Curved slopes (1, 2, 5) show PCRs in which MSO = 1 to MSO = 479 are weaker riders drafting the strongest rider (479 W). Straight and curved slopes converge at PCR points 0.1, 0.62, and 0.99 [29].

As long as $d$ is constant, inputting different MSOs for strongest rider in PCR Eq. (2) (e.g. set max MSO in a given range as 375 W), moves the slope intersection points along the range axis, but does not change the intersection point equivalence to $d$ that corresponds to the PBT [29]. While the effects of two $d$ extremes are shown (0.1, and 0.99), these are constants for all speeds; further work is required to show the effects of $d$ variations according to changing speed.

In quadrant III the sample range low (MSO = 305 W) bounds the overall sample range ~36% which is near the energy savings (38%) for $d = 0.62$. Under the framework here, the nearness of these values is not coincidental and is evidence for the hypothesis that the sample range includes cyclists whose competitive differences are equalized within the range 1-$d$, corresponding to the energy savings permitted by drafting. This implies that if the sample range was substantially broader, at maximal speed set by the strongest rider, the peloton would divide such that at least one new sub-group would form comprising MSO range corresponding to 1-$d$. However, further data from a variety of pelotons is required to confirm this hypothesis.

Conversely, if a peloton exhibits an MSO range less than the 1-$d$ equivalent, the group could accommodate additional riders whose MSOs were lower than the low end of the existing range. This is shown in quadrant IV, where the range for curve 2 extends beyond the existing sample boundary of 305 W by a small margin, to range = 0.38, before curve 2 intersects PCR = 1. This means that riders with MSOs slightly less than 305 W could fit within the sample range and sustain the speed of the strongest rider. Similarly, and more obviously, curve 5 ($d = 0.1$, 90% energy savings) extends out to range = 0.90 before it intersects with PCR = 1. All cyclists with MSOs on this curve are free-riders up to PCR = 1 when the strongest rider drives the pace at her MSO, while a proportionately small number on curve 6 would engage in passing behavior. This is a consequence of the high energy savings that permits a proportionately high number of weaker cyclists to sustain the pace of the strongest rider.

Where the energy savings is small ($d = 0.99$, 1% energy savings), there is virtually no free-riding range (curve 1): no weaker cyclists can sustain the pace of stronger cyclists who ride at their MSO. Passing behavior can occur only when stronger cyclists follow weaker cyclists (curve 3). This implies that all cyclists will separate and travel at their own maximum speeds without the equalization that drafting affords.

The following general principles are extracted:

1. Over time the range of cyclists' MSOs narrows such that it is equivalent to the energy saved by drafting (1-$d$).
2. As speeds are increased up to a maximal speed set by the strongest rider, cyclists whose MSOs fall outside (below) this range will separate from the peloton (PCR ≥ 1).

3. Cyclists can engage in cooperative behavior (passing and sharing most costly front position) in combinations of MSOs at speeds such that PCR ≤ *d*. Stronger cyclists in drafting positions will always be able to pass weaker cyclists[3].
4. Weaker cyclists can sustain the pace of the strongest rider between PCR = *d* and PCR = 1 (*d* < PCR < 1), as free-riders. In this PCR range, equivalent to the energy savings (1-*d*), these cyclists cannot pass and share the most costly front position.
5. These principles imply that if a peloton exhibits a range of MSOs greater than the energy savings equivalent of drafting (1-*d*), the peloton will divide such that sub-groups fit within the range equivalent of 1-*d*.

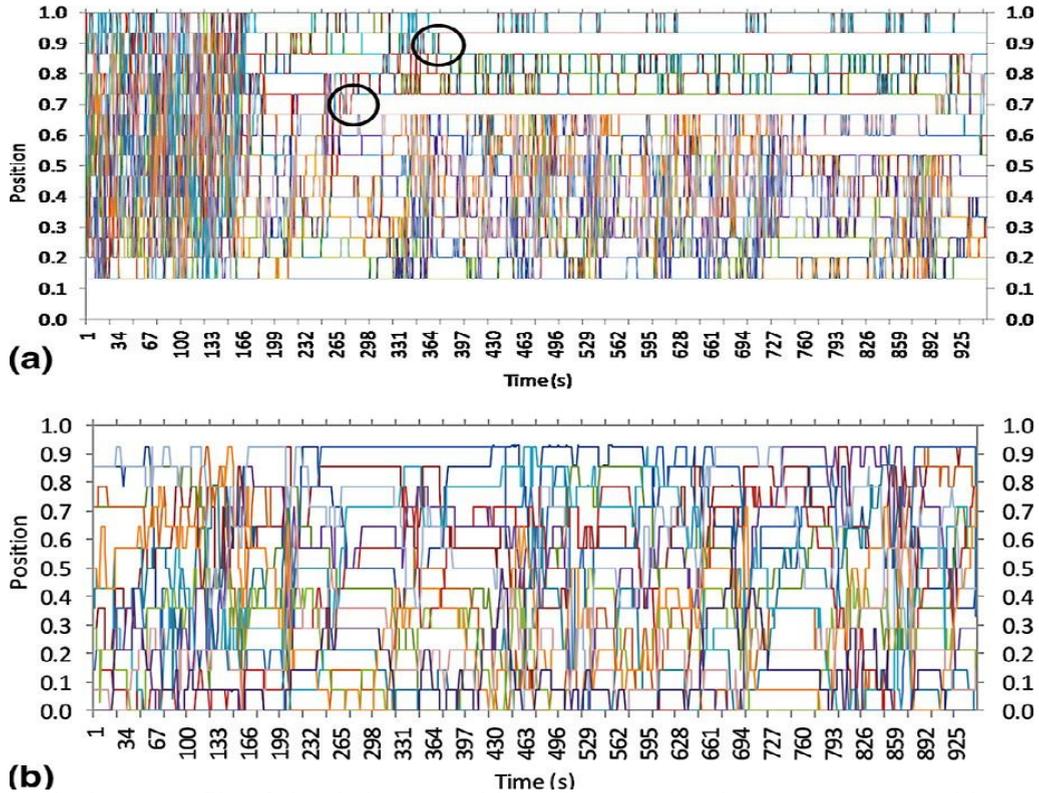

**Fig. 3.** (a) Typical race profile of simulation run with random acceleration parameter. "Position" refers to the proportion of riders ahead of the focal rider. Circles indicate peloton divisions. (b) Mass-start race position profile from [15,19].

## 2. Methods

### 2.1. Model validation

The simulation model in [15] is constructed first by applying the cohesion and separation algorithm from [19], as modified from Wilenski's flocking algorithm [30]. Eqs. (2)–(6) represent the mathematical peloton model of "backward" passing by deceleration, and are incorporated in the simulation algorithm. An adjustable peloton speed parameter (converted to power outputs in the simulation) represents the speed of the front rider. The peloton speed can be manually adjusted to trigger the range of self-organized positional dynamics as determined by the mathematical model. Simulation algorithm includes drag parameters affecting power output that assumes flat topography, windless conditions, and standard cyclists' power output parameters, as set out in [15]. The MSOs applied were the same values for 14 simulated cyclists as used in [15]. Cyclists MSOs were randomly generated between 305 W and 479 W for all simulations reported here.

---

[3] The speed at which a stronger cyclist can pass a weaker one diminishes as their MSOs approach equality, but this factor is not modelled here.

For simulations generated by the deceleration algorithm (Eqs. (3)–(6)), changes in relative position are determined by the requirement for cyclists to decelerate relative to the peloton speed and their own maximal capacities. In this respect fatigue is modeled by the threshold PCR = 1 as an instantaneous parameter such that cyclists cannot sustain speeds exceeding PCR = 1. To the extent that fatigue has longer lasting effects that produce decelerations at reduced MSOs, those are model refinements that may be the subject of future research.

The simulation model in [15] contains a weak collective coherence quantity (a flocking algorithm parameter [30]) to enable a minimum cyclist cluster density [19] and speed differential among cyclists. This coherence quantity is not based on realistic speed differentials so any resulting forward passing (by acceleration) behavior is not accurately simulated. Here we introduce a random acceleration parameter (RAP) to generate passing behavior based on more realistic random variations and differentials in acceleration speeds, but only if PCR < 1. If PCR > 1, no acceleration is possible and any RAP that would extend a simulated cyclists' current speed beyond its MSO equivalent, triggers deceleration according to Eqs. (3)–(6). Speeds set by the front rider (peloton speed) are always a tunable parameter. The addition of a RAP allows us to observe simulated cyclists' "attempts" to adjust positions forward within a realistic range of outputs up to MSOs. The RAP also allows observation of predicted differences in speeds between groups that occur partly as a function of group size [14,20], and other unknowns.

The updated model incorporating RAP 0.0–2.0 m/s was validated by comparing its simulated position profile, as shown in Fig. 3a,[4] against the actual race position profile shown in Fig. 3b[5] from [28] (mass-start race with 14 cyclists). Position profiles show each cyclist's relative position second to second, providing a complete visual representation of cyclists' changing positions for the entire race duration.

---

[4] The zero (0) position is not shown because the simulation counting algorithm is such that agents always count themselves, and so there is always at least one cyclist "ahead".

[5] In opposite fashion to footnote 3, the one (1) position is not shown because the counting procedure (counted by hand from video footage) was such that there were always cyclists ahead of the last position (i.e. if you are in last position you do not count yourself among those ahead). Ties for last position reduce the total ahead by the number of cyclists tying for the last position.

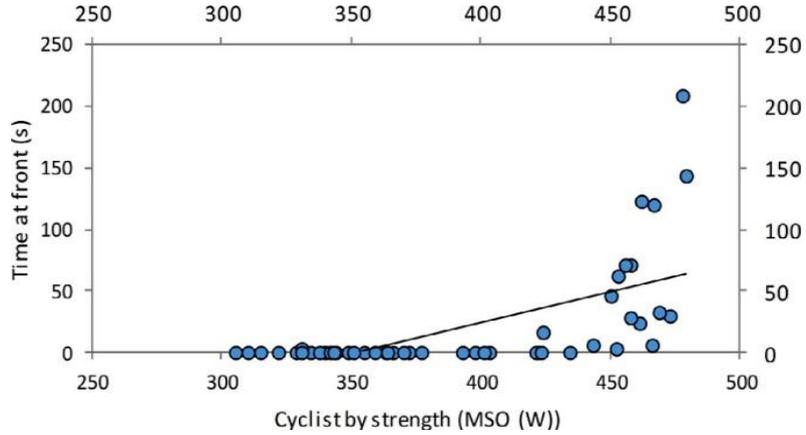

**Fig. 4.** Typical results for 50 simulated cyclists driven to collective speeds of 11.5 m/s + RAP 0.0 m/s to 2.0 m/s, showing proportionate durations in highest cost front position ($R$ = 0.7574). The strongest riders (highest MSOs) spend more time at the front, while weaker cyclists must decelerate in order to sustain outputs where PCR < 1, and so they do not rotate through the front position, becoming free-riders.

Position profiles are generated by applying Eq. (2) from [28]:

$$P = \frac{C_{ahead}}{C_{total}}/\Delta t \qquad (8)$$

Here $P$ is position; $C_{ahead}$ is the number of riders ahead, discounting riders aligned laterally (those directly beside or behind); $C_{total}$ is the total number in the peloton; $t$ is the period between observations, in seconds, here every 1 s.

One significant difference between the model version presented here and the version in [15] was an increased tendency for the group of 14 simulated cyclists to divide into groups (typically into two), as in Fig. 3a. This typical division is explained by the presence of the RAP, which permits cyclists' to accelerate within the given random range immediately after a collective acceleration without any refractory period due to fatigue. Individual refractory periods are expected after high intensity bursts, [31] and collective peloton recovery periods were identified in [28]. As discussed, this should be a model refinement for future study.

Multiple simulations were run to check similarity between the simulation profile and the actual mass-start race profile, with a typical result shown in Fig. 3a. Fig. 3a and b show reasonable similarity, indicating a sufficiently valid model for running Test Protocols A and B.

### 2.1.1. Test Protocol A

The objective of these tests was to identify the distribution of cyclists' time spent at the front of the peloton as a function of their MSOs. The hypothesis was that as peloton speeds increase, correspondingly fewer cyclists can share time at the front. Tests were run involving 5, 10, 25, 50, and 100 simulated cyclists. MSOs for each group were randomly generated between 305 W and 479 W at collective speeds 9.5 m/s, 11.5 m/s (both including RAP range 0.0–2.0 m/s), and 6 m/s with RAP range 0.0–5.5 m/s. RAP range 0.0–5.5 m/s is reasonable when peloton speeds are lower at 6 m/s, to accommodate greater latitude for cyclists to accelerate from existing speeds up to their maximums.

Five tests were run for each group of differing size (i.e. five trials with 5 cyclists, five with 10, etc.) for each of the three different speeds (5 trials × 5 groups × 3 speeds: 75 tests in total). Cyclists' durations at the front of each peloton were compared. Results are robust: runs were 1000 time steps each equivalent to 16:40 min, approximately equal to the duration of the race event studied in [28]; total 5000 time steps for each speed, the equivalent of 1.38 h of race footage for each speed.

It was expected that as speeds were increased, a sorting process would occur such that only those cyclists with the highest MSOs would sustain time spent at the front, while weaker cyclists would be forced to decelerate in order to maintain PCR < 1, either by following in drafting positions or by simply proceeding more slowly than the front rider.

### 2.1.2. Results protocol A

Results reflect the expected behavior. When simulated cyclists are driven to a peloton speed that approaches or exceeds many (but not all) cyclists' MSOs, a sorting process occurs such that the strongest riders spend more time at the front of the peloton (also shown in [15]). For example, when peloton speeds are driven to 11.5 m/s, the strongest riders sustain and exchange positions at the front because speeds do not force these particular strong riders to exceed their MSOs, while many others are forced to decelerate to maintain PCR < 1, as shown in Fig. 4. Those forced to decelerate become free-riders because they do not contribute to the most costly front position.

18 of 50 simulated cyclists spent some time (> 0 s) at the front; 17 of these were MSO ≥ 424. Cyclist with MSO = 478 spent 20.8% of the 1000 time steps at the front, MSO = 479 (14.4%), MSO = 462 (12.3%), MSO = 467 (12%). One cyclist among the 18 with an MSO < 424 (MSO = 331), spent 0.3% of the test event at the front. Of the remaining 32 simulated cyclists who spent 0 s at the front, four were MSO > 400; all others were MSO < 400. Thus there is strong correlation between cyclists' strength and their time at the front when speeds are high. This is because, at these sufficiently high speeds, only the strongest cyclists possess sufficiently high MSOs to sustain the speeds without drafting.

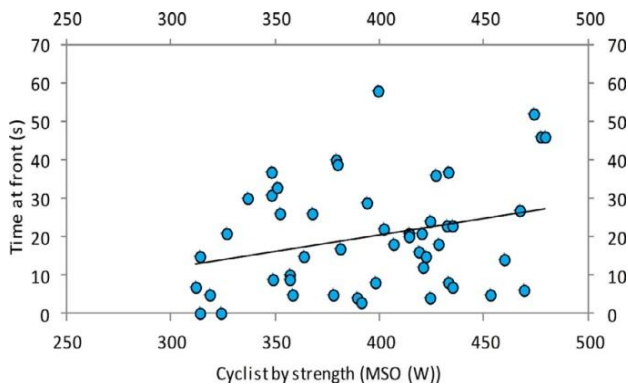

**Fig. 5.** Typical results for 50 simulated cyclists driven to collective speed from 9.5 m/s + RAP 0.0 m/s to 2.0 m/s, showing relative times spent in most costly, front position. There is minimal correlation between cyclists' strength and time spent at the front ($R$ = 0.2835), indicating high degree of random mixing and passing within the peloton.

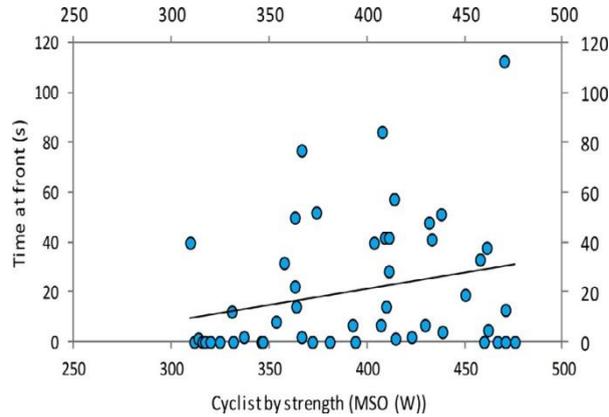

**Fig. 6.** Typical results for 50 simulated cyclists driven to collective speed from 6.5 m/s + RAP 0.0 m/s to 5.5 m/s, showing relative times spent in most costly, front position ($R$ = 0.2546). Due to the wide range of random accelerations, we expect high passing, mixing, and sharing of front position, similar to results in Fig. 5.

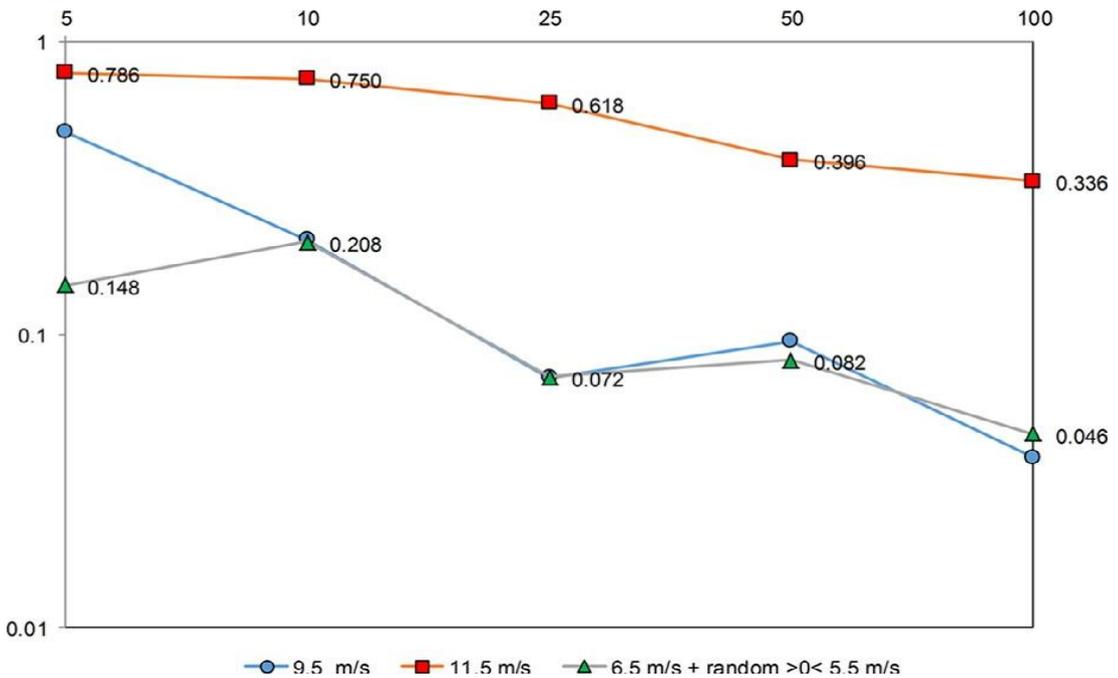

**Fig. 7.** $R$ values as a function of group size for tests at three speeds (total 75 tests). Differences in $R$ values between tests at 11.5 m/s compared with tests at 9.5 m/s and 6.5 m/s + random acceleration 0.0–5.5 m/s are clear for groups of size > 5.

By comparison, when speeds are sufficiently low for all group members to travel below their MSOs, there is minimal correlation between cyclists' strength and the time they spend at the front of the peloton, as shown in Figs. 5 and 6. As shown in Fig. 5, the percentage of cyclists' time spent at the front is not well correlated to strength. 48 of 50 cyclists spent some time at the front (>0 s), ranging from 3 s to 58 s. One cyclist with MSO = 399 spent 5.8% of its time at the front, the greatest duration. The next three in hierarchy of percentages – MSOs = 474 (5.2%), 479 (4.6%), and 477 (4.6%) – are of high MSO, while the sample balance is roughly randomly distributed for time at the front. The distribution of cyclists' time at front is similar for results shown in Fig. 7.

These results confirm the expectation that cyclists comfortably – in terms of energy costs – share time spent at the front of the peloton at relatively low speeds, and this time distribution is more equitable than when collective speeds are higher. Further, the random equitable distribution is indicated by $R$ values that are markedly lower for groups of size 10, 25, 50, 100 than when peloton speeds are comparatively high (11.5 m/s), as in Figs. 4–6.

These results discount behavior in actual bicycle races in which organized teams may deliberately control the pace and dominate time at the front at comparatively low speeds; i.e. these results confirm natural self-organized dynamics traceable to basic physical/physiological principles, which are likely to be observed in non-human biological systems.

Additional, similar results for pelotons composed of 5, 10, 25, and 100 cyclists, are summarized in terms of $R$ values for times spent at front, as shown in Fig. 7, below. Similar to Fig. 4, the comparatively high $R$ values for groups travelling at high collective speeds (11.5 m/s) are consistent among peloton groups of different size. Similar to Figs. 5 and 6, $R$ values are comparatively low for lower peloton speeds, indicating general sharing of cyclists' time spent in the most costly front position. For small groups of 5 at lower collective speeds, results lack consistency and are not considered reliable.

### 2.1.3. Test Protocol B

The objective of these tests was to induce divisions within the peloton to observe: 1. new group formations; 2. how groups sort hierarchically according to strength (group mean MSO); and 3. evidence of protocooperative behavior consistent with the stated principles. The protocooperative behavior theory here implies that, after peloton divisions, sub-groups should exhibit PCRs that approach equivalent $d$. The MSO range was constant for all tests: 305–479 W (range = 0.36) to reflect the range of MSOs among the set of cyclists competing in the mass-start race referred to in [15,28]. Future research should test broader ranges for their sorting results and the PCRs of new groups.

Ten tests of 50 simulated cyclists were run in which cyclists were timed for 1000 s (16:40 min, roughly the duration of the mass-start race in [28], generating the equivalent of 2.78 h of race footage). As in [15], individual MSOs were generated randomly, distributed between 305 W and 479 W. The relationships between MSO, group size, speeds, and PCRs were analyzed.

Cyclists were initially activated at peloton speed 9.5 m/s (34.2 km/h)(with RAP 0.0–2.0 m/s) for 15 s, a sufficiently low speed to enable high-density clustering and passing behavior. Cyclists were then accelerated to peloton speed 12.5 m/s (45 km/h) (with RAP 0.0–2.0 m/s) for 30 s, and then decelerated to peloton speed 10 m/s (36 km/h) (with RAP 0.0–2.0 m/s) for the remainder of the test event. 30 s accelerations were used to be consistent with standard sprint test durations [31], sufficient to induce peloton divisions. 12.5 m/s (45 km/h) was selected as the peloton acceleration speed since the maximum cyclist MSO was 479 W, equal to maximum non-drafting sustainable speed of 13.2 m/s (47.53 km/h)[6]. Thus a peloton acceleration speed of 12.5 m/s (45 km/h) ensured the majority of cyclists were driven to their MSO (PCR ≥ 1), sufficient to cause realistic peloton divisions within 30 s of this high speed. This also accounts for time required for weaker cyclists who may be positioned near the front of the peloton to shift backward through the peloton and separate, although this factor is not quantified in this analysis.

## *2.14. Results Test Protocol B*

Ten test runs generated a total of 36 groups of varying size between 2 and 48 cyclists. These groups self-organized as a result of divisions and merges induced by accelerations for 30 s at speeds of 12.5 m/s (45 km/h). A group was identified as being separate from another if there was at least 6 *x*-coordinate positions (1 *x*-coordinate = 1 m) between cyclists of greatest separation. After the 30 s acceleration period, typically between two and four sub-groups formed, which subsequently either merged or divided again, such that groups reconstituted in size and number over the test duration. Frequently, once a division occurred, groups continued to diverge at different mean speeds.

Typically, divisions and merges occurred in the first half of the test event, after which relative group positions stabilized until the finish. A typical test run profile is shown in Fig. 8. This form of continual divergence and reintegration is expected in actual competitive pelotons [14,15,28].

There is negative correlation between mean group speeds and mean PCR. Generally, for the speed ranges tested, as mean group speeds fall, their mean PCRs increase, as in Fig. 9a. This is somewhat surprising since we might expect slower groups to have lower mean PCR. However, it is explainable by the lower mean MSO in slower groups, as shown in Fig. 9b, which drives PCRs higher on average for slower groups approaching their maximums at the given peloton speed. Speeds would need to be substantially lower to see corresponding drops in PCR for these groups.

Generally, group speed is independent of group size, with some tendency for speeds to increase with the size of the group, as in Fig. 10a. This tendency is consistent with the findings of [14,20]. The low correlation here reflects situations in which small groups would, for some duration, travel faster than larger groups. PCRs tend to fall with group size, as in Fig. 10b. This indicates increased drafting probability (free-riding) as group size increases.

A relatively strong correlation is observed between group mean PCRs and their mean MSOs, as shown in Fig. 11a. At equal or nearly equal speeds, groups with lower mean MSO will travel at higher PCR than groups with higher mean MSO, consistent with
Fig. 9a.

There are 11 of 36 data points (groups) of PCR < 0.62, which is some evidence of groups exhibiting generally cooperative behavior. The nearness of these 11 group means to PCR = 0.62 (PCR = 0.581–0.614) has at least two possible explanations: 1. it is a consequence of the range 305–479 W (range = 0.36, close to 1-*d*) and the mean speeds in which sub-groups tend to be driven close to their maximums, since at lower speeds we might expect to see lower mean PCRs; 2. PCR < 0.62 indicates weak riders ahead of stronger ones (when speeds driven to MSO), where PCR > 0.62 indicates strong riders ahead of weak ones (when speeds driven to MSO) and a mean PCR approaching 0.62 indicates passing behavior and an equitable changing of positions at speeds approaching MSO, but not at MSO. This is evidence of cooperative behavior. The greater number of groups (25/36) having PCR > 0.62 indicates a tendency toward predominantly free-riding behavior, consistent with [13]: a result that predicts similar behavior in other biological contexts.

---

[6] Determined from www.bikecalculator.com; rider weight = 64 kg, bicycle weight = 8 kg, clincher tires, drops position, 0% grade, 0 headwind, 25 degrees C, 18 m elevation.

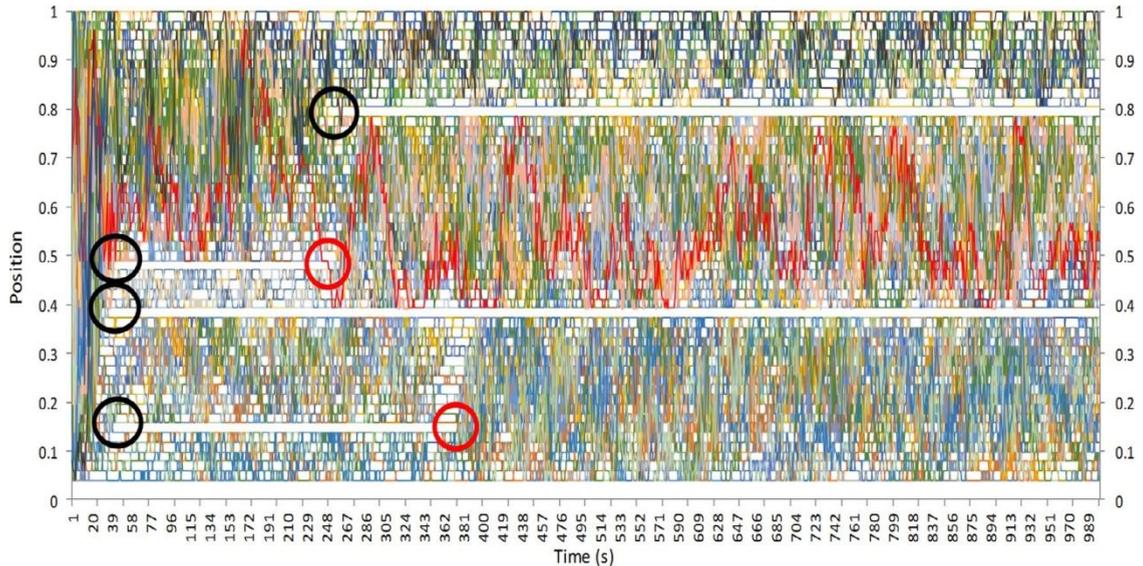

**Fig. 8.** Profile of simulated peloton, 50 cyclists (1000 s), showing division and reintegration points. Collective acceleration at time 15 s, sustained for 30 s. Position value 0 indicates the front position (not computed[3]), while value 1 shows rider at the back of the peloton (Eq. (6)). Black circles indicate division points, while red circles indicate merge points. Four groups emerged initially (~45 s), a merge and a division occurred at ~250 s, and a merge occurred at ~375 s, leaving three groups for the last ~2/3 of the test. Profiles indicate high mixing or positional change in the middle group, and fewer positional changes at front and rear of the peloton. Vertical slope intersections indicate passing; horizontal parallel slopes indicate group stretching and reduced passing.

## 3. Discussion and directions for future work

Results for Test Protocol A support the proposition that cooperative behavior occurs when groups travel at mean MSOs below a critical PCR threshold, which we show is equivalent to *d* (~PCR = 0.62), as in Figs. 1 and 2. Above that threshold, cyclists are incapable of passing, but with aid of drafting may still maintain the speed of a rider ahead, up to a de-coupling threshold (PCR = 1). This is the free-riding range.

Results for Test Protocol B, in simulated sorting conditions, shows 11/36 groups PCR < 0.62 (cooperative range), and 25/36 groups PCR > 0.62 (the free-riding range). Because results PCR < 0.62 are near PCR = 0.62, the simulated sorting results are somewhat ambiguous in terms of the hypothesized cooperative behavior below PCR = 0.62, and data was not obtained of the durations of relative times at the front. However, PCR > 0.62 implies that strong riders are more often in front of weak riders as speeds approach maximums, which is consistent with Fig. 4. This implies further the presence of mixed cooperative and freeriding behavior, if stronger riders tend to circulate near the front while weaker riders free-ride. However, although it is clear there is considerable passing behavior within each group, as shown in Fig. 8, it is unknown the degree to which subsets of stronger riders shared front-riding positions for the groups in Test Protocol B, as shown in Fig. 4.

Future work involving self-organized sorting as in Test Protocol B should isolate mixed-phase behavior within a given group *before* divisions occur to show how subsets of narrower MSO ranges emerge within larger cohesive groups. Future work should also involve tests with expanded ranges through a greater range of speeds (particularly lower speeds), and with variable $d$.

Another avenue for further work is the effect of fatigue after group division, and the degree to which fatigue may cause further group deceleration or delay reintegration between groups. Test results do not reflect situations of group capitulation, when strong cooperators fatigue and groups relax, decelerating considerably relative to other groups. In such situations, PCRs would correspondingly fall and we would expect cooperative behavior to increase thereafter. Further work could include a fatigue parameter to explore its effect on PCR and protocooperative behavior.

## 4. The peloton superorganism

The concept of sports comprising a superorganism has been proposed [32], but to our knowledge no rigorous academic treatments exist of a peloton as a superorganism. In addition to the mechanics of protocooperative behavior, the self-organized process of divisions and merges characterizes a competitive process because of the implicit group objective to obtain resources ahead of others. It is the tension, or "tug-of-war" between intra-group cooperative behavior and inter-group competition that characterizes the peloton as a superorganism, as the criteria is generally stated in [18]. This tension in the context of bicycle racing was mentioned in [33], without rigorous discussion. Here we do not rigorously apply the model parameters in [18], but rather identify in principle some analogous features that characterize a peloton as a superorganism.

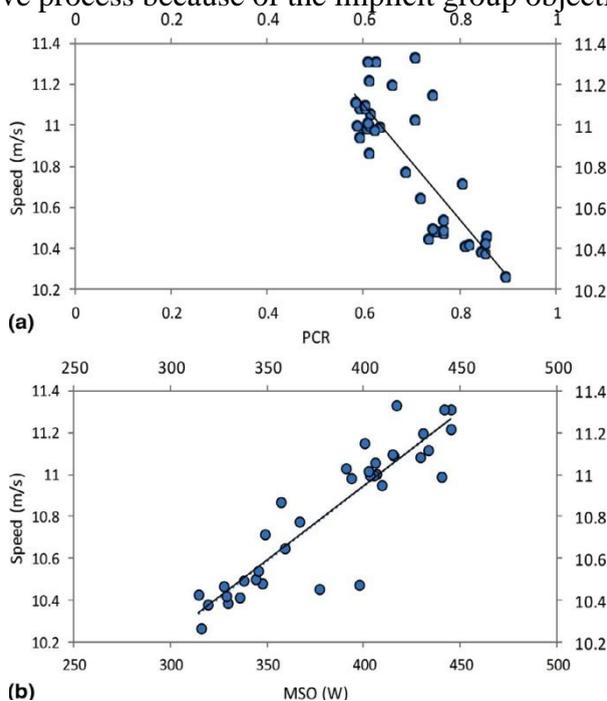

**Fig. 9.** (a) Typical result showing comparatively strong negative correlation between group mean speed and PCR ($R = -0.832$). (b) Typical result showing strong correlation between mean peloton speed and MSO. Each plot value represents a group varying in size between 2 and 48 cyclists ($R = 0.908$).

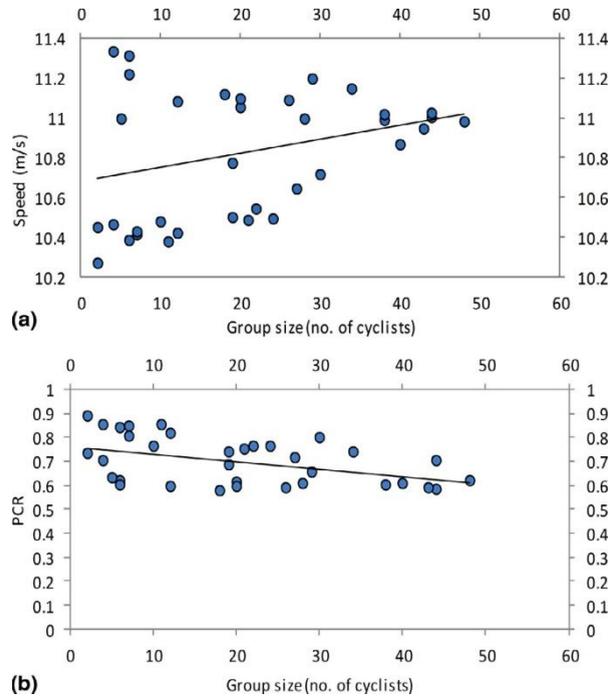

**Fig. 10.** (a) Typical result showing minimal correlation between mean speeds and peloton size. Line of best fit indicates weak tendency for speeds to increase with group size ($R = 0.297$). (b) Typical result showing correlation between mean group PCR and group size. Weak correlation indicates that PCR tends to fall as group size increases ($R = 0.449$).

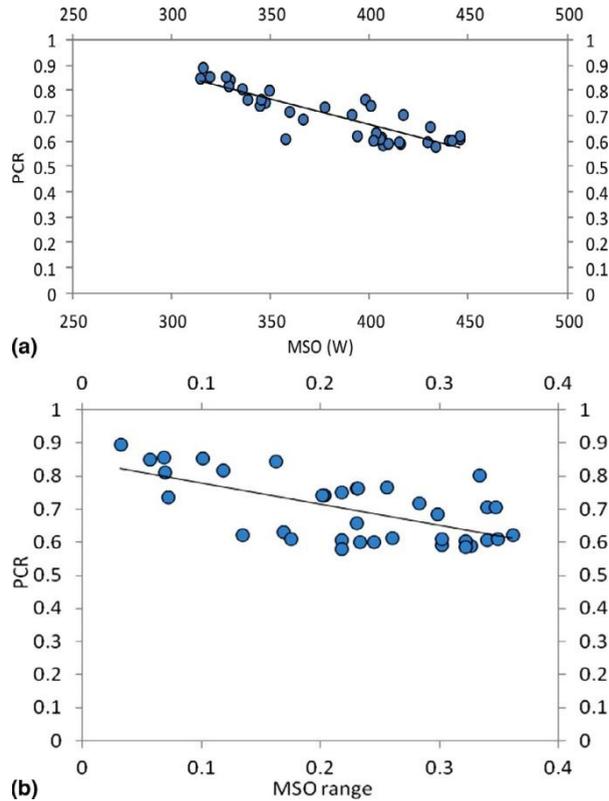

**Fig. 11.** (a) Typical result showing correlation between group mean PCR and MSO ($R = -0.862$). PCRs cluster between ~0.58 and ~0.90 (11 data points PCR < 0.62). (b) Correlation between mean group PCR and MSO ranges for each group ($R = 0.628$). Where group MSO ranges are narrower than 1-$d$ (i.e. < 0.38), PCRs are greater than $d$ and that as group range approaches 1-$d$ (0.38), PCR approaches $d$ (0.62). Here no ranges are greater than $d$, so there is no PCR data for ranges > $d$ equivalent.

The conclusions in [18] are that an individual's evolutionary stable investment in within-group cooperation versus within group competition:

a. increases as within-group relatedness increases;
b. decrease as group size increases;
c. increases as the number of competing groups in a patch increases;
d. decreases as between-group relatedness increases.

The peloton model indicates that intra-group cooperation is a function of each individual's current output as a ratio of their maximum physiological capacity, and is not connected to relatedness. However, we may think of relatedness in terms of narrow ranges of maximum capacities, and groups with narrow ranges tend to exhibit increased cooperative behavior at sufficiently low relative speeds. This is compared to more widely heterogeneous groups (generally expected in larger groups) in which mixed phase behavior is more likely, and when stronger individuals will cooperate while weaker individuals tend to free-ride, as shown in Fig. 4. This is consistent with criterion "b" above. Further, the larger the group, the greater the probability that cyclists spend more time on average in drafting positions, and not sharing the most costly front-position, as implied by Fig. 10b.

In this view and consistent with the concept of protocooperative behavior, consideration of ranges of maximal capacities instead of relatedness may provide insight into a more primitive stage of evolution that precedes the emergence of genetic relatedness and cooperative behavior among kin.

In cases of wide heterogeneity by strength (as expected in the case of high population groups), we may expect the group to divide into sub-groups whose ranges are narrower after the division, increasing cooperative behavior for stronger groups, and decreasing it for weaker groups at the same collective speeds. This is consistent with criterion "a" above for groups below the protocooperative behavior threshold, but inconsistent for those groups above the PBT (i.e. as long as speeds are sufficiently low for the given group, criterion "a" is satisfied).

Similarly, peloton divisions that result in multiple sub-groups is consistent with criterion "c", particularly for stronger groups, but less so for weaker groups at the same collective speeds. As the number of groups increases, so does the probability that any given individual will share the most costly front position.

In terms of criterion "d", where a widely heterogeneous peloton divides into sub-groups, the closer groups are to each other in strength, the more likely each will sustain their maximum speeds: a following group to catch a group ahead, and the leading group to stay ahead of the following group. Any group travelling at near maximal speed involves increased free-riding and reliance on the strongest members to maintain their respective apparent objectives (catch group ahead, or stay ahead of group behind). This results in decreased intra-group cooperation.

This implies that the closer groups are in strength, the greater the probability they will maintain their apparent objectives; the greater the disparity, the less likely weaker groups will continue chasing stronger ones. While consistent with criterion "d", this is supported more by practical observation than by the model here. The model here accounts for decreased intra-group cooperation for weaker groups generally above-PBT at a given peloton speed, while stronger groups exhibit increased cooperation below PBT at the same given peloton speed. A model that introduces a fatigue factor could more closely demonstrate criterion "d", since at a given peloton speed, weaker groups will likely weaken and decelerate faster than stronger groups, thus altering relative within-group cooperation.

## 6. Conclusion

A theoretical framework for protocooperative behavior in pelotons is proposed. A mechanism determining a threshold for self-organized cooperative and free-riding behavior is modeled in bicycle pelotons. Because this behavior is driven by basic non-volitional energetic and physiological principles, we refer to it as protocooperative behavior. The protocooperative behavior threshold and the free-riding range are hypothesized to be present in any biological system in which there is an energy savings mechanism. Further, we show the tension between intra-group cooperation and inter-group competition is largely consistent with characteristics of a superorganism.

The model is well supported by Test Protocol A; and supported by some results of Test Protocol B, but inconclusive in other aspects.

These are main principles and conclusions:

- Protocooperative behavior emerges in pelotons as a function of coupled energetic costs in which there is an energy savings mechanism (drafting) for following cyclists.

- The result is a tendency to share the most costly non-drafting positions at comparably low speeds when cyclists' capacity to pass others is abundant. This tendency diminishes as collective speeds increase and free-riding increases.

- The protocooperative behavior threshold is equivalent to the coefficient of drafting $d$; below this threshold cooperative passing behavior and sharing of costly front positions occurs, while above it free-riding is a physiological necessity.

- When groups are driven to their maximal sustainable capacities as set by the strongest member(s), sub-groups form such that their ranges of MSOs are equivalent to the energy savings of drafting ($1-d$); or less if a sub-group is too small to attain MSO range equivalent to $1-d$.

In the evolutionary context, these principles have important implications for the scaling of group mean capacities, and how and where group niches form and speciation occurs.

Findings are consistent with the conclusion that cooperation emerges at lower speeds. How collective mobility and speed affects cooperation is a comparatively recent extension of cooperation theory, whereby there is some agreement that lower speed mobility contributes to cooperation more than high speed mobility [34–37].

Among other parallels, peloton dynamics represent a migration problem in terms of energy savings mechanisms and coupled dynamics, similar to migratory birds [38] or schools of fish [39–41]. Advances of cooperation theory into the domain of short and long-term migration [42,43], seem well benefitted by peloton theory. Further, theory here provides a basis for investigations of sorting among groups when being chased at near maximal speeds by predators and how differential ranges of metabolic capacities in species occurs.

Although we have studied the effects of increasing speeds and energetic demands on cooperative behavior, we have not studied the effects of fatigue periods which may delay peloton divisions or induce them more rapidly at some reduced MSO. Further, fatigue induces delays in the resumption of passing behavior after decelerations have occurred. A refined model and simulation should account for individual and collective refractory periods before cyclists resume accelerations. Similarly, a refined model could include considerations of time required for weaker cyclists to shift backward through a peloton when these cyclists are driven to unsustainable speeds; this time duration depends on peloton length and cyclist speed differentials [28].

Generally, further work is required to fully develop the theoretical framework of protocooperative behavior. In these contexts and in view of the breadth of the framework proposed, some of the test results shown here should be reproduced and their implications further explored both in the context of pelotons and other biological collectives.